\documentclass[11pt,a4paper,psamsfonts]{amsproc}

\usepackage{amsmath,amsthm}
\usepackage{amssymb}
\usepackage{amscd}
\usepackage{enumerate}
\usepackage{graphicx,tikz}
\usepackage{anysize}
\marginsize{3.5cm}{3.5cm}{2.5cm}{2.5cm}

\newtheorem{theorem}{Theorem}

	\title{A 0.502$\cdot$MaxCut Approximation\\ using Quadratic Programming}
\author{Stefan Steinerberger}
\address{Department of Mathematics, University of Washington, Seattle, WA 98195}
\thanks{The author is partially supported by NSF (DMS-2123224) and the Alfred P. Sloan Foundation.}
\date\empty

\begin{document}

\begin{abstract}
We study the \textsc{MaxCut} problem for graphs $G=(V,E)$. The problem is NP-hard, there are two main approximation algorithms with theoretical guarantees: (1) the Goemans \& Williamson algorithm uses semi-definite programming to provide a $0.878 \cdot \textsc{MaxCut}$ approximation (which, if the Unique Games Conjecture is true, is the best that can be done in polynomial time) and (2) Trevisan proposed an algorithm using spectral graph theory from which a $0.614 \cdot \textsc{MaxCut}$ approximation can be obtained. We discuss a new approach using a specific quadratic program and prove that its solution can be used to obtain at least a $0.502 \cdot\textsc{MaxCut}$ approximation. The algorithm seems to perform well in practice.\end{abstract}

\maketitle

\section{Introduction and Statement}

\subsection{Introduction} The \textsc{MaxCut} problem for a given graph $G=(V,E)$ asks for a partition $V = A \cup B$ such the number of edges that run between the sets $A$ and $B$ is maximized
$$ \textsc{MaxCut}(G) = \max_{A,B \subset V \atop A \cap B = \emptyset} \# E(A,B).$$
The problem is NP-hard, even  approximating \textsc{Max-Cut} by any factor better than $16/17 \sim 0.941$ is NP-hard \cite{bell, hastad}. 
We recall the simplest randomized algorithm: by putting each vertex randomly into one of two sets and taking the expectation, we see that
$ \textsc{MaxCut}(G) \geq 0.5 \cdot |E|.$

\begin{center}
\begin{figure}[h!]
\begin{tikzpicture}[scale=1]
\filldraw (0,0) circle (0.06cm);
\filldraw (-0.5,1) circle (0.06cm);
\filldraw (0.5,2) circle (0.06cm);
\filldraw (0.5,1) circle (0.06cm);
\filldraw (3,0) circle (0.06cm);
\filldraw (3.2-0.5,1.1) circle (0.06cm);
\filldraw (3.7,1.8) circle (0.06cm);
\filldraw (3.5,1.2) circle (0.06cm);
\draw [thick] (0,1) ellipse (1cm and 1.6cm);
\draw [thick] (3.2,1) ellipse (1cm and 1.6cm);
\draw [thick] (0.8,2) -- (2.3,1.7);
\draw [thick] (1,1) -- (2.2,1.5);
\draw [thick] (0.8,0) -- (2.2,0.5);
\draw [thick] (1,0.5) -- (2.2,0.3);
\draw [thick] (1,1) -- (2.2,1.1);
\node at (-1.5, 1) {$A$};
\node at (4.5, 1) {$B$};
\end{tikzpicture}
\caption{Finding a partition $V= A \cup B$ with $\# E(A,B)$ large.}
\end{figure}
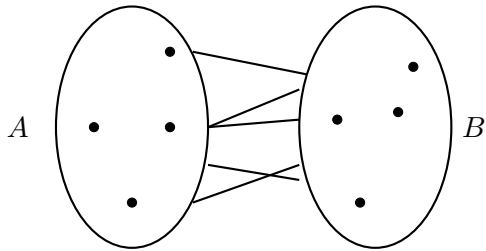
\end{center}
No algorithm improving on this elementary estimate was known until Goemans \& Williamson \cite{goemans} introduced their seminal algorithm:
they relax 
$$ 2 \cdot |E| - 4 \cdot \textsc{MaxCut}(G) =  \min_{x_i \in \left\{-1,1\right\}} \sum_{i,j=1}^{n} a_{ij} x_i x_j$$
by replacing the $x_i \in \left\{-1,1\right\}$ with unit vectors $v_i \in \mathbb{R}^n$ and $x_i x_j$ with $\left\langle v_i, v_j\right\rangle$. This is a more general problem
but one that can be solved with semi-definite programming in polynomial time. A random hyperplane cut is used to induce a partition which can be shown to be at least $0.878 \cdot \textsc{MaxCut}(G)$ in expectation. If the Unique Games Conjecture \cite{khot} is true, this is the best possible approximation ratio for \textsc{Max-Cut} that can be computed in polynomial time.
 Given the fundamental nature of the question, it would naturally be interesting to have alternative points of view. Trevisan \cite{trev2}  introduced an algorithm based on spectral graph theory: while spectral graph theory has been successful in finding minimal cuts, Trevisan shows that there is a natural dual interpretation (`anti-Cheeger-inequalities'). This algorithm provides at least a $0.531 \cdot \textsc{MaxCut}$ approximation. Soto \cite{soto} improved the constant to $0.614$.

\subsection{Related results.} 
For practical considerations, Burer, Monteiro \& Zhang \cite{burer3} suggested a relaxation of the Goemans-Williamson approach which works very well in practice \cite{dunning}. Related algorithms were recently given by the author \cite{stein}. These types
of algorithms present themselves with a curiously benign energy landscape which is of intrinsic interest \cite{boumal2, ling, stein2}. They also have a certain similarity with Kuramoto oscillators \cite{burer3, kuramoto, jianfeng, taylor} and recent hardware-based oscillator approaches to \textsc{MaxCut}, see \cite{chou, mallick, wang, wang2}.
Regarding specifically the use of quadratic programming for \textsc{MaxCut}, we refer to Charikar \& Wirth \cite{charikar} and to \cite{chan, de, hopkins, kothari} for work on linear programming.

\section{The Result}
\subsection{Main Result}
Let $G=(V,E)$ be a given graph with $|V| = n$ vertices. We use $D \in \mathbb{R}^{n \times n}$ to denote the diagonal matrix containing the degrees of the vertices, i.e. $D_{ii} = \deg(v_i)$. $A \in \left\{0,1\right\}^{n \times n}$ denotes the adjacency matrix, i.e. $A_{ij} = 1$ if $v_i$ and $v_j$ are connected and 0 otherwise. We can assume that the graph is simple and thus $A_{ii} = 0$. We propose the following explicit quadratic program
$$
(QP) = \begin{cases}
\min_{x \in \mathbb{R}^n} \qquad &\left\langle x, D^{-1} A D^{-1} x \right\rangle \\
\mbox{subject to}~& 0 \leq x_i \leq \deg(v_i)\\
\mbox{and} & \sum_{i=1}^{n}x_i \geq |E|.
\end{cases}
$$
 $D$ and $D^{-1}$ are diagonal matrices, $A$ is symmetric and thus so is $D^{-1} A D^{-1}$.
The restrictions always admit a nonempty feasibility set: the handshake lemma
$$ \sum_{v \in V} \deg(v) = 2 |E|$$
shows that the vector
$$x = \left(\frac{\deg(v_1)}{2}, \frac{\deg(v_2)}{2}, \dots, \frac{\deg(v_n)}{2}\right) \in \mathbb{R}^{n \times n}$$
 satisfies all conditions. We note that $D^{-1}AD^{-1}$ will usually not be positive definite.
\begin{theorem} Let $x \in \mathbb{R}^n$ be a solution of $(QP)$ and 
$C = \left\{v \in V: x_{v} \geq 0.23 \cdot \deg(v) \right\}.$
 If $\textsc{MaxCut}(G) \geq 0.995 \cdot |E|$, then
 $$ \# E(C, V \setminus C) \geq 0.502 \cdot \textsc{MaxCut}(G).$$
\end{theorem}
If $\textsc{MaxCut}(G) \leq 0.995 \cdot |E|$, then assigning vertices randomly to two sets gives a cut of size $0.5\cdot |E| \geq 0.502 \cdot  \textsc{MaxCut}(G)$
and thus we can always get a cut of that size.
There is no indication that this argument is sharp in terms of constants. 
The proof does not require $x$ to be a minimizer, it only requires that the quadratic form is sufficiently close to the minimum (which could be relevant in practice).

\subsection{The Idea.}
The underlying idea, which perhaps also has other manifestations, is as follows: we think of the vertices as little devices that hold a certain amount of $L^1-$mass (with $v \in V$ holding a total of $x_{v}$).
Each vertex can only hold an amount of mass that is at most as large as the degree of the vertex: vertices with more neighbors can hold more mass. At the
same time, the condition
$$ \sum_{i=1}^{n}x_i \geq |E|$$
forces that a substantial amount of mass is present in the system.

\begin{center}
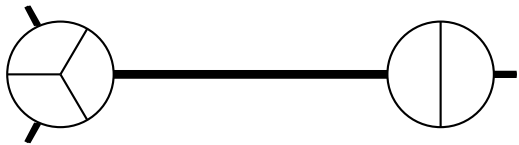
\begin{figure}[h!]
\begin{tikzpicture}[scale=1]
\draw [thick] (0,0) circle (0.7cm);
\draw [thick] (5,0) circle (0.7cm);
\draw [ultra thick] (0.7, 0) -- (4.3, 0);
\draw [ultra thick] (0.7, 0.03) -- (4.3, 0.03);
\draw [ultra thick] (0.7, -0.03) -- (4.3, -0.03);
\draw [thick] (0,0) -- (0.35, 0.6);
\draw [thick] (0,0) -- (0.35, -0.6);
\draw [thick] (0,0) -- (-0.7, 0);
\draw [ultra thick] (-0.3, 0.65) -- (-0.43, 0.9);
\draw [ultra thick] (-0.28, 0.65) -- (-0.42, 0.9);
\draw [ultra thick] (-0.32, 0.65) -- (-0.45, 0.9);
\draw [ultra thick] (-0.3, -0.65) -- (-0.43, -0.9);
\draw [ultra thick] (-0.28, -0.65) -- (-0.42, -0.9);
\draw [ultra thick] (-0.32, -0.65) -- (-0.45, -0.9);
\draw [thick] (5, -0.7) -- (5,0.7);
\draw [ultra thick] (5.7, 0) -- (6,0);
\draw [ultra thick] (5.7, 0.02) -- (6,0.02);
\draw [ultra thick] (5.7, -0.02) -- (6,-0.02);
\end{tikzpicture}
\caption{Thinking of vertices as holding probability mass.}
\end{figure}
\end{center}
\vspace{-20pt}
We can then rewrite and interpret the quadratic form
$$ \left\langle x, D^{-1} A D^{-1} x \right\rangle = \left\langle D^{-1}x, A D^{-1} x \right\rangle$$
in a new way. $D^{-1} x$ is merely the vector describing the percentages of maximal charge that the vertices are charged with. Abbreviating this by $p = D^{-1}x$, the
inner product can thus be written as
\begin{align*}
  \left\langle D^{-1}x, A D^{-1} x \right\rangle &= \sum_{i,j=1}^{n} a_{ij} p_i p_j = 2 \sum_{e \in E \atop v \sim_e w} p_{v} p_{w}.
\end{align*}
The entire system has to hold a certain amount of charge, so not all the $p_v$ can be small. Minimizing the quadratic form thus amounts
to making sure that vertices with a large amount of charge are mainly surrounded by vertices with a small amount of charge. This then
suggests that the vertices with a large amount of charge should have relatively few edges running between them and might be a good
candidate for a partition. A crucial factor is that at least some vertices holding a large amount of charge are necessarily
adjacent to many edges -- if they were adjacent to few, their capacity would be too small. This idea seems quite general and can perhaps also be implemented using other types of quadratic programs.

\subsection{In practice.} One would naturally wonder how the algorithm does in practice. We used the following simple implementation: solve
$(QP)$, define the sets
$$ A = \left\{ v \in V: \frac{x_v}{\deg(v)} \geq \frac12\right\} \qquad \mbox{and} \qquad  B = \left\{ v \in V: \frac{x_v}{\deg(v)} < \frac12\right\}$$
and use $\#E(A,B)$ as a lower bound for $\textsc{MaxCut}(G)$.
We used the \textsc{QuadraticOptimization} command in Mathematica 12 for computations. When comparing to the Goemans-Williamson algorithm,
we find that the two methods are broadly comparable in terms of performance on Erd\H{o}s-Renyi random graphs (see Table 1).
\begin{center}
\begin{table}[h!]
\begin{tabular}{ l c c  }
  Graph & QP  &  GW \\
  \hline
  $G(50, 0.3)$ & 236 & 234 \\
$G(50, 0.5)$  & 368 & 363 \\
$G(100, 0.1)$  & 327  & 343  \\
$G(100, 0.5)$  & 1399  & 1398  \\
$G(200, 0.1)$  & 1281  & 1260  \\
\end{tabular}
\vspace{10pt}
\caption{Average size of cuts obtained by (QP) and Goemans-Williamson over 1000 random Erd\H{o}s-Renyi graphs $G(n,p)$.}
\end{table}
\end{center}

\vspace{-10pt}

There are many variations on both methods that one could consider (simply in terms of solvers, solutions heuristics, etc.). For example, since $D^{-1}AD^{-1}$ and $x$ only have nonnegative entries, the condition $ \sum_{i=1}^{n}x_i \geq |E|$ could also be replaced by $\sum_{i=1}^{n}x_i = |E|$.
Regarding variations for
$(QP)$, we mention
$$
(QP)_{\alpha, \beta} = \begin{cases}
\min_{x \in \mathbb{R}^n} \qquad &\left\langle x, D^{-1} A D^{-1} x \right\rangle \\
\mbox{subject to}~& 0 \leq x_i \leq  \alpha \deg(v_i)\\
\mbox{and} & \sum_{i=1}^{n}x_i \geq \beta|E|.
\end{cases}
$$
The approach discussed above uses $\alpha = \beta =1$.
We found, empirically, that it may sometimes be advantageous to work with slightly different values of $\alpha, \beta$. We also observed that $x_{v}/\deg(v)$ tends to be quite close to 0 or 1 for most vertices with rather few values in
between: if there are many `ambiguous' cases, this may indicative of not having yet found the true solution and in that case
 slightly adjusting $\alpha, \beta$ may help to shift towards `better' solutions.

\section{Proof of the Theorem}
We assume, throughout the argument, that $x \in \mathbb{R}^n$ is a solution of 
$$
(QP) = \begin{cases}
\min_{x \in \mathbb{R}^n} \qquad &\left\langle x, D^{-1} A D^{-1} x \right\rangle \\
\mbox{subject to}~& 0 \leq x_i \leq \deg(v_i)\\
\mbox{and} & \sum_{i=1}^{n}x_i \geq |E|.
\end{cases}
$$
We first show that if $\textsc{MaxCut}(G) \sim |E|$, then $\left\langle x, D^{-1} A D^{-1} x \right\rangle$ is small.
In the second step we show that if $\left\langle x, D^{-1} A D^{-1} x \right\rangle$ is small, then taking the set $C$ of vertices 
for which $x_v \geq \eta \deg(v)$ (with the parameter $\eta$ eventually chosen to be $\eta \sim 0.23$) leads to $\# E(C, V \setminus C)$
improving on the trivial bound.\\

\textbf{1. Restricting to $\varepsilon$ small.}
Let us assume that
$$ \textsc{MaxCut}(G) = (1 - \varepsilon) |E|.$$
We know that $\varepsilon \leq 0.5$. 
We can always get a cut of size $0.5 \cdot |E|$ which shows that it is always possible to get an
\begin{align*}
 \mbox{cut of size} &\geq \frac{0.5 \cdot |E|}{ (1 - \varepsilon) |E|}\cdot \textsc{MaxCut}(G)\\
&=  \frac{1}{2-2\varepsilon}\cdot \textsc{MaxCut}(G).
\end{align*}
This is the trivial argument and naturally gets close to $1/2$ as soon as $\varepsilon$ becomes small. We now assume that $\varepsilon$ is small and argue using the
quadratic program.\\

\textbf{2. If $\varepsilon$ is small, the solution of $(QP)$ is small.}
We take a \textsc{MaxCut}-solution and call the partitions $U$ and $W$. Then, by construction,
$$ \# E(U,W) = \textsc{MaxCut}(G) =  (1 - \varepsilon) |E|.$$
This implies that
$$ \#E (U,U) + \# E(W, W) \leq \varepsilon |E|.$$
Since $V = U \cup W$
$$ \sum_{v \in U} \deg(v) + \sum_{v \in W} \deg(v) = \sum_{v \in V} \deg(v) = 2 |E|,$$
we know that for at least one of the two sets, the sum is at least $|E|$. We assume without loss of generality that this set is $U$, i.e. that
$$ \sum_{v \in U} \deg(v) \geq |E|.$$
We will show that $(QP)$ has a small solution by showing that there exists a feasible vector leading to a small value of the quadratic form.
We choose the test vector $y \in \mathbb{R}^n$ to be $y_v = \deg(v) 1_{U}(v)$,
where $1_U$ is the indicator function of $U$. The vector $y \in \mathbb{R}^n$ satisfies all constrains of $(QP)$ and
\begin{align*}
 \left\langle y, D^{-1}A D^{-1}y \right\rangle &=  \left\langle D^{-1}y,A D^{-1}y \right\rangle = \left\langle 1_{U}, A ~1_{U} \right\rangle  =  2~\#E(U,U) \leq 2\varepsilon |E|.
 \end{align*}
This shows that when $\textsc{MaxCut}(G) =  (1 - \varepsilon) |E|$ and if $x$ is a solution of $(QP)$, then
$$ \left\langle x, D^{-1} A D^{-1} x \right\rangle \leq 2 \varepsilon |E|.$$

\textbf{3. If $(QP)$ has a small solution, we get a large cut.}
Let us now conversely suppose that $x \in \mathbb{R}^n$ satisfies all the constraints and
$$ \left\langle x, D^{-1}A D^{-1}x \right\rangle \leq 2\varepsilon |E|.$$
We define, for $0 < \eta < 1$ (to be determined later), the set
$$ C = \left\{v \in V: x_v \geq \eta \cdot \deg(v) \right\}.$$
Using the handshake lemma and the constraints on the variables,
\begin{align*}
|E| &\leq \sum_{v \in V} x_v = \sum_{v \in C} x_v +  \sum_{v \in V\setminus C} x_v\\
&\leq \sum_{v \in C} \deg(v) + \sum_{v \in V\setminus C} \eta \cdot \deg(v) \\
&=(1-\eta) \sum_{v \in C} \deg(v)+ \eta \sum_{v \in C} \deg(v) + \sum_{v \in V\setminus C} \eta \cdot \deg(v) \\
&= (1-\eta) \sum_{v \in C} \deg(v) + 2\eta |E| .
\end{align*}
This implies that there are many edges that connect to a vertex in $C$ 
$$ \sum_{v \in C} \deg(v) \geq \frac{1 - 2\eta}{1-\eta} \cdot |E|.$$
We also observe that, by positivity of all involved quantities,
\begin{align*}
2\varepsilon |E| &\geq  \left\langle x, D^{-1}A D^{-1}x \right\rangle\\
&=   \left\langle D^{-1}x,A D^{-1}x \right\rangle \\
&\geq  \left\langle (\eta 1_{C}), A  ~(\eta 1_{C}) \right\rangle = \eta^2 \left\langle  1_{C}, A~ 1_{C} \right\rangle = \eta^2 \cdot 2  \#E(C,C).
\end{align*}
and therefore
$$ \#E(C,C) \leq  \frac{\varepsilon}{\eta^2} |E|.$$
Using $\deg_C(v)$ and $\deg_{V \setminus C}(v)$ to denote the number of neighbors a vertex $v \in V$ has in the sets $C$ and $V \setminus C$, respectively, we see that
\begin{align*}
\frac{1 - 2\eta}{1-\eta} \cdot |E| &\leq \sum_{v \in C} \deg(v) \\
&= \sum_{v \in C} \deg_{C}(v) + \deg_{V \setminus C}(v) \\
&= 2 \#E(C,C) + \#E(C, V \setminus C).
\end{align*}
From this, we can conclude that, using the inequality above and the definition of $\varepsilon$,
\begin{align*}
 \# E(C, V \setminus C) &\geq \left( \frac{1 - 2\eta}{1-\eta} - \frac{2\varepsilon}{\eta^2}\right) |E| \\
 &= \left( \frac{1 - 2\eta}{1-\eta} - \frac{2\varepsilon}{\eta^2}\right) \frac{1}{1 - \varepsilon}\textsc{MaxCut}(G).
 \end{align*}
 
 \textbf{4. Optimizing $\eta$.}
This means we can always find a cut (either probabilistically or via the quadratic program) of size at least
$$ \# E(C, V \setminus C) \geq \max\left\{  \frac{1}{2-2\varepsilon},  \left( \frac{1 - 2\eta}{1-\eta}- \frac{2\varepsilon}{\eta^2}\right)\frac{1}{ 1-\varepsilon}\right\} \textsc{MaxCut}(G).$$
Some quick algebra shows that the best constant arises from setting
$$ \eta = \frac{1}{6}(5-\sqrt{13}) \sim 0.232\dots$$
For this choice of $\eta$, we obtain
$$\min_{0 \leq \varepsilon \leq \frac{1}{2}} \max\left\{ \frac{1}{2-2\varepsilon},  \left( \frac{1 - 2\eta}{1-\eta} - \frac{2\varepsilon}{\eta^2}\right) \frac{1}{1-\varepsilon}\right\} =  \frac{1}{139} \left(23+13 \sqrt{13}\right) \sim 0.502\dots$$

\end{document}